\documentclass{osa-article}

\journal{osajournal}

\usepackage{xcolor}

\begin{document}

\title{Optical Angular Momentum manipulations in a Four Wave Mixing process}

\author{Nikunj Prajapati,\authormark{1,*} Nathan Super,\authormark{1} R. Nicholas Lanning,\authormark{2} Jonathan P. Dowling,\authormark{2,3,4,5} and Irina Novikova\authormark{1}}

\address{\authormark{1}Department of Physics, The College of William and Mary, Williamsburg, VA 23185, USA}
\address{\authormark{2} Hearne Institute for Theoretical Physics and Department of Physics $\&$ Astronomy, Louisiana State University, Baton Rouge, Louisiana 70803, USA}
\address{\authormark{3} CAS-Alibaba Quantum Computing Laboratory, USTC, Shanghai 201315, China}
\address{\authormark{4} NYU-ECNU Institute of Physics at NYU Shanghai, Shanghai 200062, China}
\address{\authormark{4} National Institute of Information and Communications Technology, Tokyo 184-8795, Japan}
\email{\authormark{*}nprajapati@email.wm.edu} 



\begin{abstract}
We investigate the spatial and quantum intensity correlations between the probe and Stokes optical fields produced via four-wave mixing in a  double-$\Lambda$ configuration, when both incoming probe and control fields carry non-zero optical orbital angular momentum (OAM). We observed that the topological charge of the generated Stokes field obeyed the OAM conservation law. However, the maximum values and optimal conditions for the intensity squeezing between the probe and Stokes fields were largely independent of the angular momenta of the beams, even when these two fields had significantly different OAM charges. We also investigated the case of a composite-vortex pump field, containing two closely-positioned optical vortices, and showed that the generated Stokes field carried the OAM corresponding to the total topological charge of the pump field, further expanding the range of possible OAM manipulation techniques.  
\end{abstract}

\section{Introduction}
Structured light --- an optical beam carrying optical angular momentum (OAM) --- rapidly became a useful resource for a wide range of applications, from optomechanical manipulations~\cite{Knoner:07,Ilice1602738} to super-resolution imaging~\cite{OAM_imaging_via_focus,Allen_L_1992}. In quantum information science it has been used for the generation of hyper-entanglement~\cite{Lui_oam_2014}, quantum multiplexing~\cite{Bozinovic_2013}, etc. 
Nonlinear optical processes offer effective methods for OAM manipulations~\cite{marino_oam_2008, Walker_oam_transfer_2012, Akulshin:16, Kyohhei_2013, Lassen_2009}, as the OAM phase-matching conditions make it possible to control the spatial structure of the generated optical fields by shaping the profiles of strong pump and weak probe fields before the interaction. 


Four-wave mixing in hot Rb vapor has provided a valuable method for producing spatially multi-mode squeezed and entangled optical fields~\cite{Jing_2017,marino_oam_2008,Lassen_2009}. 
For example, a four-wave mixing interaction between a strong pump and a weak probe in a double-$\Lambda$ configuration, shown in Fig.~\ref{fig:setup}, results in the amplification of the probe field, and the generation of the Stokes field.
Previous studies have shown that OAM that is present in the initial control or probe field will determine the spatial mode of the generated Stokes field while maintaining a high degree of quantum intensity  correlations between the probe and Stokes fields~\cite{lettPRA08}. 
The focus of our experiment is to extend the parameter space of accessible OAM modes by independently implanting OAM on both the pump and probe optical fields. 
We demonstrate that, despite considerable differences in spatial profiles of the generated probe and Stokes fields, we observed comparable levels of quantum intensity correlations. 

\begin{figure}[ht!]
\centering\includegraphics[width=0.7\columnwidth]{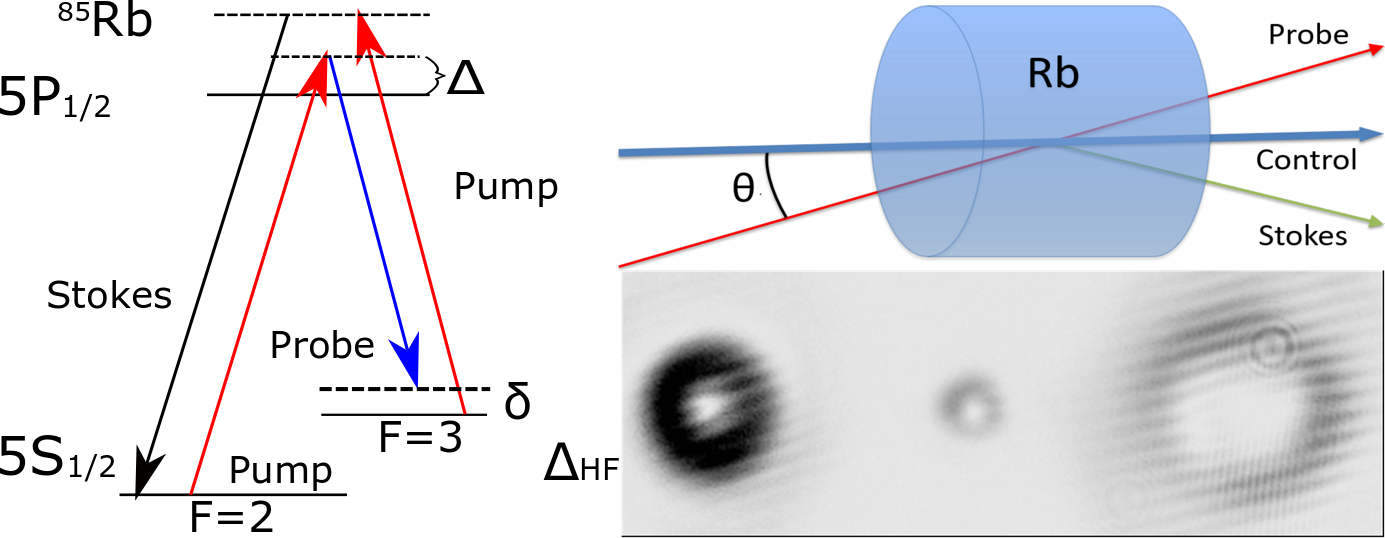}
\caption{Realization of the four-wave mixing in ${}^{85}$Rb vapor: \emph{(a)} level diagram and  \emph{(b)} geometrical arrangement. \emph{(c)} Example of the output probe and generated stokes field with mutual difference of $\Delta \ell = 4$. }\label{fig:FWM_levels}
\end{figure}

To achieve high four-wave mixing gain and a high level of intensity correlations between the probe and stokes fields, the phase matching conditions $\vec{k}_{\mathrm{probe}} + \vec{k}_{\mathrm{Stokes}} = 2 \vec{k}_{\mathrm{pump}}$ must be obeyed, forcing the two input optical fields to cross inside the Rb cell at the proper angle, with the Stokes field being then generated symmetrically with respect to the output pump beam. 

If either input optical field carries optical angular momentum, the OAM phase-matching condition dictates the topological charge of the generated Stokes field:

\begin{equation} \label{Eq:OAMconservation}
\ell_{\mathrm{Stokes}} = 2 \ell_{\mathrm{pump}} - \ell_{\mathrm{probe}}.
\end{equation}  

Thus, it should be possible to produce the Stokes field carrying OAM in a significantly broader range than if only one of the input beams carried OAM~\cite{marino_oam_2008}. 
As an example, Fig.~\ref{fig:FWM_levels}(c) shows the Stokes field with $\ell_{\mathrm{Stokes}}=-3$ that is generated using probe and pump fields carrying the unit topological charges of the opposite sign ( $\ell_{\mathrm{probe}}=+1$ and  $\ell_{\mathrm{pump}} =-1$). 
This results in the topological charge difference of $\Delta \ell = 4$ between the two quantum correlated optical fields. 
Yet, as we will show in the following sections, such manipulations of the spatial beam profiles do not cause significant deterioration of the quantum correlations between the two fields.
Moreover, we found that the four-wave mixing process allows for the use of an optical field with a composite vortex structure to increase the total effective topological charge of a beam. 
Specifically, we used a phase mask, which added two closely-separated, but clearly distinguishable, unit charge vortices to the input pump field. 
By analyzing the generated Stokes field profile, we unambiguously demonstrate that the Stokes field OAM value is consistent with the total OAM carried by the pump field, rather than with that of an individual vortices. 
These observations suggest that FWM can be used as a mechanism for effectively merging separate topological defects, thus realizing a new tool for OAM manipulations.
  
\section{Experimental arrangements}

\begin{figure}[ht!]
\centering\includegraphics[width=1\columnwidth]{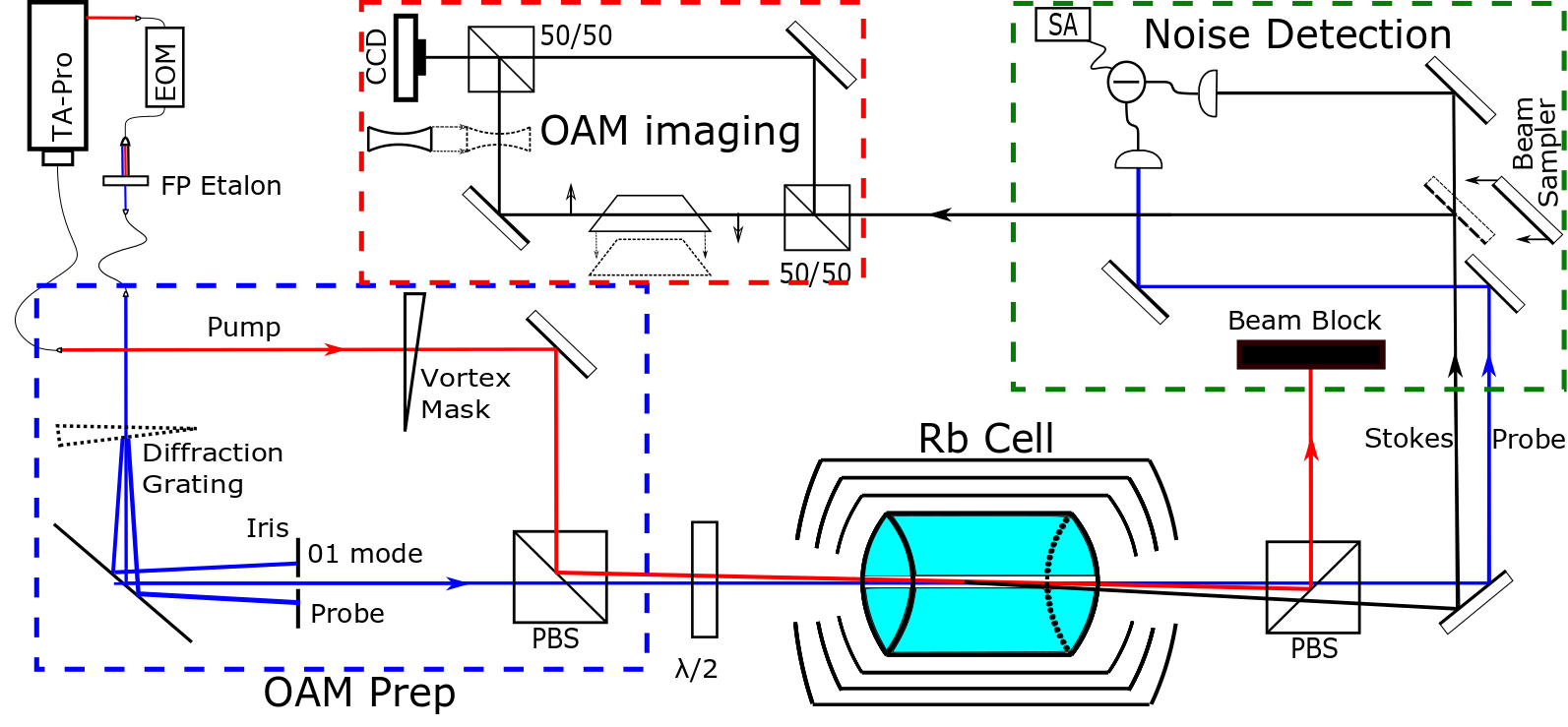}
\caption{The optical layout of the experimental setup. A single laser is used to generate all outputs. The high-power pump (red) is output though a fiber dock system while the lower power pump output is in free space and later used to generate the probe (blue). The stokes (black) optical field is generated through FWM in the cell. The set-up has four main sections; probe prep, OAM prep, interaction, noise detection, and imaging. Abbreviations and explanations are given in the text.}\label{fig:setup}
\end{figure}

The schematic of the experimental setup is shown in Fig.~\ref{fig:setup}. To maintain phase coherence, the pump and probe were taken from the same diode laser tuned to 1.25 GHz to the blue of the $5S\_1/2,F=2->5P\_1/2,F'=2$ optical transition of the ${}^{85}$Rb atoms. A portion of the laser output was sent to a tapered amplifier to generate the pump field. The rest was phase-modulated at the frequency $3035$~MHz (the hyperfine splitting of $^{85}$Rb) using a fiber electro-optical modulator. The low-frequency modulation sideband, tuned to 1.25 GHz to the blue of the $5S\_1/2,F=3->5P\_1/2,F'=2$ transition, served as a probe field (the other modulation sidebands were filtered out using a Fabry-Perot (FP) etalon). Both the probe and pump field then passed through single-mode fibers to ensure their Gaussian transverse intensity profiles.
We then used two different methods to control the topological charges of the two input optical fields. For the probe field, we used a forked diffraction grating that directed $\approx 50\%$ of the input intensity into the first diffraction maximum, thus preparing the probe field with the  spatial charge of $\ell=1$. For the pump field, we used a transparent spiral vortex phase mask to add $\ell=\pm 1$ OAM charge, depending on the mask orientation without significant optical power losses.  

The OAM-carrying pump and probe optical fields were then combined at a proper phase-matching angle ($\approx 0.4^\circ$) using a polarizing beam splitter (PBS) before entering a 25-mm long Pirex cell filled with isotopically enriched $^{85}$Rb vapor. The cell was mounted inside of a three-layer magnetic shielding and maintained at ~$106^\circ$C corresponding to the atomic density of $7\times 10^{12}~\mathrm{cm}^{-3}$. At the location of the cell, the pump and probe beams had diameters of 250~$\mu$m and 300~$\mu$m and powers of 410~mW and 60~$\mu$W, respectively. After the cell, the pump beam was filtered out using a second PBS, and the amplified probe and generated Stokes fields were spatially separated using an edge mirror and sent to the two inputs of a balanced photodetector for the differential intensity measurements.

To analyze the vortex structure of the output beams, we deployed two interference methods. 
In the first one, either the probe or Stokes beam was individually passed through a Mach-Zehnder interferometer with a divergent lens placed in one of its arms, such that at the output, the original vortex beam overlapped with a constant phase section of the expanded beam. 
Their interference pattern produced a traditional forked interferogram. 
This method allowed us to easily identify the position(s) and number of vortices in the original beam by simply counting the number of forked fringes in the resulting interferogram. 
For a more accurate analysis of the OAM beam conposition, we alternatively replaced the lens with a Dove prism in one of the arms that transposed the beam. 
The interference of the original and the transposed optical fields resulted in a petal interferogram, in which the azimuthal phase difference between the beams in two interferometer arms produced a flower-like structure with the number of petals equal to the double of the input beam topological charge~\cite{spiralbandwidthpreprint}.

\section{Optical angular momentum conversion with single vortex beams}

In the first series of measurements, we independently prepared both the pump and probe optical fields in pure LG modes with unit topological charge, as described above. 
During all of the measurements, the probe was kept in the same ($\ell_{\mathrm{probe}} = +1$) LG mode. 
However, by flipping the orientation of the phase mask, we set the topological charge of the pump field to be either $\ell_{\mathrm{pump}} = \pm1$. 
Two configurations were tested: when the pump and probe optical fields had the same ($\ell_{\mathrm{probe}}=\ell_{\mathrm{pump}}=1$) or opposite ($\ell_{\mathrm{probe}}=-\ell_{\mathrm{pump}}=1$) OAM charges. 
For each configuration, our goal was to test the OAM phase matching in Eq.~(\ref{fig:FWM_levels}) by analyzing the intensity and phase profile of all optical fields after the cell.
At the same time, we measured the intensity correlations between the output probe and stokes fields to confirm that the intensity squeezing is preserved, even if the two fields are in different transverse modes.

\begin{figure}[ht!]
	\centering\includegraphics[width=1\columnwidth]{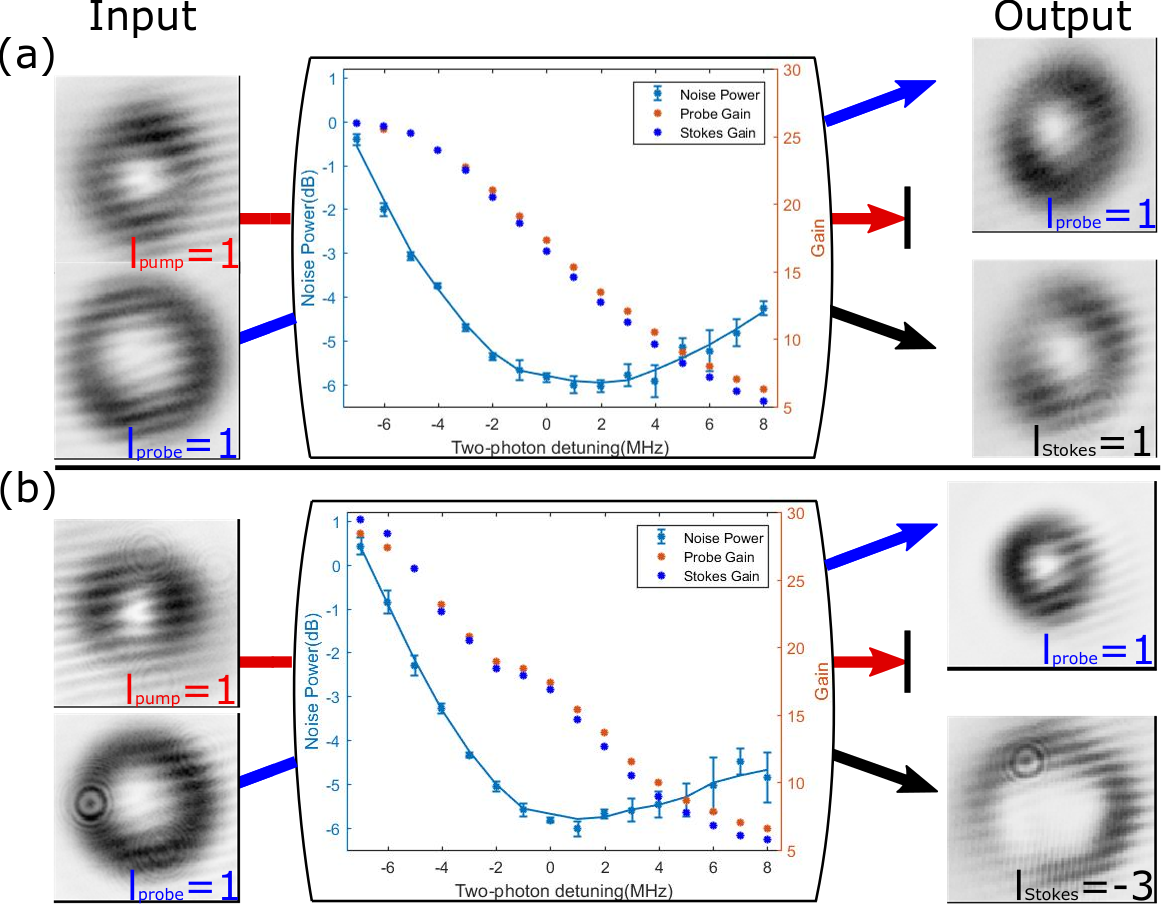}
	\caption{\emph{Left:} Intensity squeezing (left axis) and the FWM amplification for the probe and Stokes fields (right axis) as functions of the two-photon detuning, measured  for (a) $\ell_{\mathrm{probe}}=+1$, $\ell_{\mathrm{pump}}=+1$, and (b) $\ell_{\mathrm{probe}}=+1$, $\ell_{\mathrm{pump}}=-1$ configurations. Images on the left show the flat-front interferograms of the input pump and probe fields at the cell's position, and on the right --- the interferograms of the output probe and Stokes fields for each configuration. }
	\label{fig:purevortices}
\end{figure}

In the first configuration, the identical topological charges in the probe and pump fields $\ell_{\mathrm{probe}}=\ell_{\mathrm{pump}}=1$ resulted in the Stokes beam being generated in the same mode,  $\ell_{\mathrm{Stokes}}=1$, in Fig.~\ref{fig:purevortices}(a) (right). The identical unit charge for both amplified probe and the generated stokes field was confirmed by the interferogram: when interfered with the plane wave, one clear fork in the interference fringes was observed for both beams.
When the phase mask in the pump field was reversed, the pump beam was implanted with a negatively charged vortex and the Stokes was generated in the $\ell_{\mathrm{Stokes}}=-3$ mode, shown in Fig.~\ref{fig:purevortices}(b)(right). This observation was in excellent agreement with the OAM phase-matching condition.

To achieve the maximum intensity squeezing, great care had to be taken to adjust the waists and the convergence of the pump and probe fields to increase their spatial overlap inside the Rb cell. The input pump and probe beams are shown on the left of Fig.~\ref{fig:purevortices}(a) and \ref{fig:purevortices}(b). In these configurations, we saw substantial FWM gain and were also able to maintain a large two-mode intensity squeezing ($-5.8 \pm 0.1$~dB) whether the pump and probe beams carried the same or opposite charge [see Figs.~\ref{fig:purevortices}(a) and \ref{fig:purevortices}(b)]. 
This value was within the standard error of our measurements when comparing to squeezed beams without OAM.

The dependence of the measured gain and two-mode intensity squeezing, on the two-photon detuning (between the probe and pump fields), is shown in Figs.~\ref{fig:purevortices}(a) and \ref{fig:purevortices}(b). 
It is similar to the previously reported measurements with conical beams~\cite{Jing_2017}. 
We defined gain, for both probe and Stokes fields, as the ratio of the output intensity to the intensity of the input probe field. 
One can see that the FWM gain peak is shifted from the exact hyperfine splitting frequency, due to the power broadening.  
At the same time, the best quantum-noise suppression occurs not at the maximum gain frequency, but on its wing closer to the two-photon resonance.

The two-photon detuning also affects the transverse profile of the output fields. 
Under the conditions for best squeezing, the intensity profiles of both the Stokes and probe fields most closely resembles those expected from a pure LG mode. 
However, closer to the region of maximum gain, the intensity distributions become uneven. 
We observe that the portion of the output beam that is closer to the pump beam emerges as more amplified. 
If the two-photon detuning is moved toward positive two-photon detuning, the gain region shift away from the pump amplifying the outer parts of the probe and Stokes fields.

\section{Optical angular momentum conversion for a composite vortex pump field}
\begin{figure}[ht!]
	\centering\includegraphics[width=.7\columnwidth]{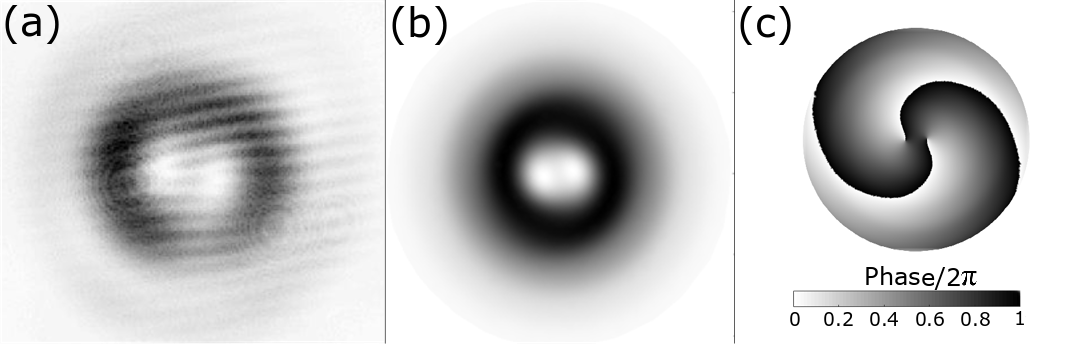}
	\caption{Composite vortex pump beam: (a) experimentally measured interferogram of the pump beam at the location of the Rb cell, (b) simulated pump beam transverse intensity distribution, described by Eq.~(\ref{Eq:compvorttheory}), and (c) a phase map of the simulated pump field.}
	\label{fig:pump}
\end{figure}

In the next series of experiments, we inserted a different phase mask, containing two closely-positioned spiral features~\cite{Maleev:03}, which produced a pump beam with two spatially separated vortices of charge $\ell =1$, as shown in Fig.~\ref{fig:pump}(a). 
To theoretically analyze the LG mode decomposition of such a beam, we assumed that the center of a spiral feature was located at $(r_0, \phi_0)$ with respect to the beam axis. 
Then we could express the $\phi$ coordinate of the phase mask as $\phi^{\prime} = \arctan (y^{\prime} / x^{\prime} ) $, where $x^{\prime} = r \cos \phi - r_0 \cos \phi_0$ and $y^{\prime} = r \sin \phi - r_0 \sin \phi_0$.
With this in mind, we modeled the modified pump beam as       
\begin{equation}\label{Eq:compvorttheory}
u_{\mathrm{pump}}(\boldsymbol{r}) = \sum c_{l,p} u_{l,p}(\boldsymbol{r}),
\end{equation}
where 
\begin{equation}
c_{l,p} = \int r dr d\phi \, u_{0,0}(r,\phi,z_{\mathrm{mask}}) \, e^{i \phi^{\prime}_1} \, e^{i \phi^{\prime}_2},
\end{equation}
and the $\phi^{\prime}_i$ coordinates correspond to the two spiral features. 
The intensity and phase distributions are shown in Figs.~\ref{fig:pump}(b) and \ref{fig:pump}(c), respectively. 
As one can see, the experimentally observed pump intensity distribution matched the theoretical one quite accurately.

\begin{figure}[ht!]
	\centering\includegraphics[width=.8\columnwidth]{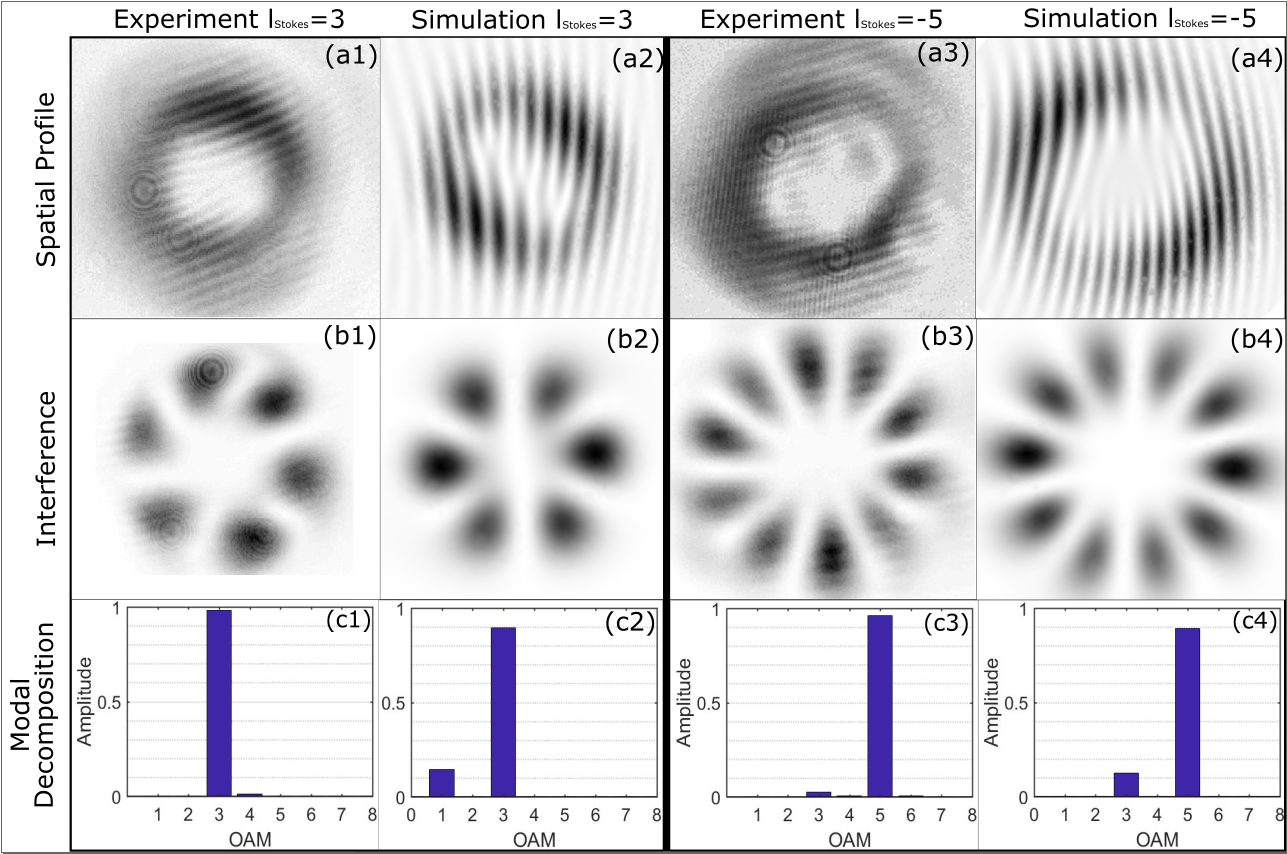}
	\caption{Top row: (a1, a3) experimentally measured and (a2, a4) numerically simulated intensity profiles of the generated Stokes field for the composite pump field, containing two $\ell=\pm1$ spatially-separated optical vortices, correspondingly. Middle row: spiral interferograms of each beam. Bottom row: Fourier mode decomposition of the radial intensity ditributions of the spectral interferograms for different LG mode indices $\ell$.}
	\label{fig:Stokes_analysis}
\end{figure}

It has been shown that such composite vortex beams can be decomposed into a superposition of pure LG modes, which allow the generation of high-dimentional entangled states~\cite{PhysRevLett.88.013601,PhysRevA.67.052313}. 
Thus, our goal was to study how the composite nature of the pump beam topological charge and the OAM conservation affect the structure of the generated stokes field. 
For example, according to Eq.(\ref{Eq:OAMconservation}), two individual pump vortices with $\ell_{\mathrm{pump}}=\pm1$ should result in the generation of a stokes field with a similar composite vortex structure, containing two vortices of either $\ell=1$ or $\ell=-3$, depending on the mask orientation. 
Thus, when the total Stokes topological charge is measured, we would expect it to be either 2 or $-6$. 
However, this is not what we observed experimentally. 
For one orientation of the phase mask,  the Stokes optical field was produced in the $\ell=3$ mode, seen in Fig.~\ref{fig:Stokes_analysis}(a1). 
In the second configuration, we observed the Stokes field generated in the $\ell=-5$ mode, seen in Fig.~\ref{fig:Stokes_analysis}(a3). 
Such behavior is consistent with the pump field contributing its total topological charge into the four-wave mixing phase matching conditions, thus behaving as a simple beam carrying $\ell_{\mathrm{pump}}=\pm2$ OAM. 
The corresponding theoretical simulations, shown in Figs.~\ref{fig:Stokes_analysis}(b2) and \ref{fig:Stokes_analysis}(b4) confirm this observation. 

The modal analysis seen in Ref.~\cite{spiralbandwidthpreprint} allowed us to more precisely quantify the distribution of $\ell$ values in the generated Stokes fields. 
Using the spectral interferograms produced in the interferometer with the inserted Dove prism [see Figs.~\ref{fig:Stokes_analysis}(b1)--\ref{fig:Stokes_analysis}(b4)], we carried out the Fourier analysis of the azimuthal intensity distribution and confirmed that the observed petal structure consist of mainly either $\ell=3$ or $\ell=-5$ LG mode with over a $90\%$ confidence, both experimentally and numerically [Figs.~\ref{fig:Stokes_analysis}(c1)--\ref{fig:Stokes_analysis}(c4)].


\begin{figure}[ht!]
	\centering\includegraphics[width=1\columnwidth]{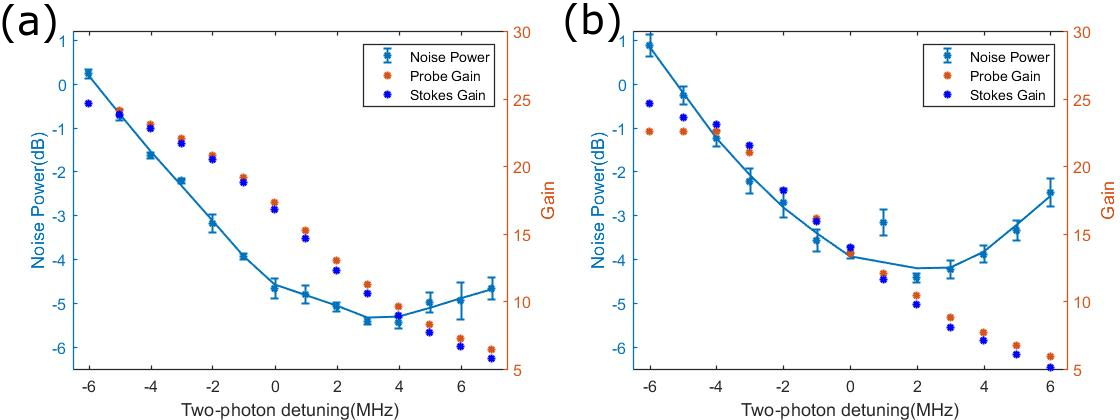}
	\centering\caption{Measured intensity squeezing (left axis) and the FWM amplification for the probe and stokes fields (right axis) as functions of the two-photon detuning, measured  for \emph{(a)} $\ell_{\mathrm{probe}}=+1$, composite pump $\ell_{\mathrm{pump}}=2\times 1$, and \emph{(b)} $\ell_{\mathrm{probe}}=+1$, $\ell_{\mathrm{pump}}=2\times(-1)$ configurations.}
	\label{fig:compvortStokes_sqz}
\end{figure}

As in the case of simple vortex beams, we saw that the intensity squeezing between the probe and stokes was maintained in these cases as well, and followed the same general trend. 
However, due to larger sizes of both Stokes and pump beam, the spatial filtering of the pump field was less efficient, resulting in small leakage of the pump signal into the detection scheme, thus reducing the detected squeezing. 
Nevertheless, in the case of Stokes generated in $\ell=3$, we measured up to nearly $-5$~dB of intensity difference squeezing [Fig.~\ref{fig:compvortStokes_sqz}(a)]. 
For the other mask orientation, in which the Stokes field was generated with the total topological charge of $\ell_{\mathrm{stokes}}=-5$, the measured squeezing level was worse, roughly $-4.3$~dB, mostly due to even larger Stokes beam size [see Fig.~\ref{fig:compvortStokes_sqz}(b)]. 
However, we believe that with the optimized detection geometry we should be able to regain the same amount of squeezing even for the beams with large topological charge difference, as in this case.

\section{Conclusion}

In this manuscript we have demonstrated control of the Stokes-field spatial-mode structure by means of shaping the input pump and probe fields using independent phase elements. We found that the Stokes-field OAM can be controlled in a much wider range without degrading the two-mode intensity squeezing between the amplified probe and generated Stokes field, regardless of their spatial mode mismatch. We also found that closely positioned phase singularities in the pump field can be effectively added in the four-wave mixing process, resulting in topological charge of the Stokes field being dependent on the total OAM of the pump, not the sum of two independent vortices. That opens an interesting avenue for manipulation of the complex spatially separated LG modes.

\section*{Funding}
NP, NS and IN supported by the National Science Foundation Grant No. PHY-308281. RNL and JPD would like to acknowledge support from the Air Force Office of Scientific Research, the Army Research Office, the Defense Advanced Research Projects Agency, the National Science Foundation, and the Northrop Grumman Corporation.  

\section*{Acknowledgments}

We would like to thank A. Arnold, S. Franke-Arnold and N. Davidson for useful comments and suggestions, G.~A. Swartzlander for lending us the vortex phase mask, and E.~E. Mikhailov for the help with the experiment.




\begin{thebibliography}{10}
	\newcommand{\enquote}[1]{``#1''}
	
	\bibitem{Knoner:07}
	G.~Kn\"{o}ner, S.~Parkin, T.~A. Nieminen, V.~L.~Y. Loke, N.~R. Heckenberg, and
	H.~Rubinsztein-Dunlop, \enquote{Integrated optomechanical microelements,}
	{\protect\JournalTitle{Opt. Express}} \textbf{15}, 5521--5530 (2007).
	
	\bibitem{Ilice1602738}
	O.~Ilic, I.~Kaminer, B.~Zhen, O.~D. Miller, H.~Buljan, and M.~Solja{\v
		c}i{\'c}, \enquote{Topologically enabled optical nanomotors,}
	{\protect\JournalTitle{Science Advances}} \textbf{3} (2017).
	
	\bibitem{OAM_imaging_via_focus}
	P.~Wo\ifmmode~\acute{z}\else \'{z}\fi{}niak, P.~Banzer, F.~Bouchard, E.~Karimi,
	G.~Leuchs, and R.~W. Boyd, \enquote{Tighter spots of light with superposed
		orbital-angular-momentum beams,} {\protect\JournalTitle{Phys. Rev. A}}
	\textbf{94}, 021803 (2016).
	
	\bibitem{Allen_L_1992}
	L.~Allen, M.~W. Beijersbergen, R.~J.~C. Spreeuw, and J.~P. Woerdman,
	\enquote{Orbital angular momentum of light and the transformation of
		laguerre-gaussian laser modes,} {\protect\JournalTitle{Phys. Rev. A}}
	\textbf{45}, 8185--8189 (1992).
	
	\bibitem{Lui_oam_2014}
	K.~Liu, J.~Guo, C.~Cai, S.~Guo, and J.~Gao, \enquote{Experimental generation of
		continuous-variable hyperentanglement in an optical parametric oscillator,}
	{\protect\JournalTitle{Phys. Rev. Lett.}} \textbf{113}, 170501 (2014).
	
	\bibitem{Bozinovic_2013}
	N.~Bozinovic, Y.~Yue, Y.~Ren, M.~Tur, P.~Kristensen, H.~Huang, A.~E. Willner,
	and S.~Ramachandran, \enquote{Terabit-scale orbital angular momentum mode
		division multiplexing in fibers,} {\protect\JournalTitle{Science}}
	\textbf{340}, 1545--1548 (2013).
	
	\bibitem{marino_oam_2008}
	A.~M. Marino, V.~Boyer, R.~C. Pooser, P.~D. Lett, K.~Lemons, and K.~M. Jones,
	\enquote{Delocalized correlations in twin light beams with orbital angular
		momentum,} {\protect\JournalTitle{Phys. Rev. Lett.}} \textbf{101}, 093602
	(2008).
	
	\bibitem{Walker_oam_transfer_2012}
	G.~Walker, A.~S. Arnold, and S.~Franke-Arnold, \enquote{Trans-spectral orbital
		angular momentum transfer via four-wave mixing in rb vapor,}
	{\protect\JournalTitle{Phys. Rev. Lett.}} \textbf{108}, 243601 (2012).
	
	\bibitem{Akulshin:16}
	A.~M. Akulshin, I.~Novikova, E.~E. Mikhailov, S.~A. Suslov, and R.~J. McLean,
	\enquote{Arithmetic with optical topological charges in stepwise-excited rb
		vapor,} {\protect\JournalTitle{Opt. Lett.}} \textbf{41}, 1146--1149 (2016).
	
	\bibitem{Kyohhei_2013}
	K.~Shigematsu, Y.~Toda, K.~Yamane, and R.~Morita, \enquote{Orbital angular
		momentum spectral dynamics of gan excitons excited by optical vortices,}
	{\protect\JournalTitle{Japanese Journal of Applied Physics}} \textbf{52},
	08JL08 (2013).
	
	\bibitem{Lassen_2009}
	M.~Lassen, G.~Leuchs, and U.~L. Andersen, \enquote{Continuous variable
		entanglement and squeezing of orbital angular momentum states,}
	{\protect\JournalTitle{Phys. Rev. Lett.}} \textbf{102}, 163602 (2009).
	
	\bibitem{Jing_2017}
	J.~Du, L.~Cao, K.~Zhang, and J.~Jing, \enquote{Experimental observation of
		multi-spatial-mode quantum correlations in four-wave mixing with a conical
		pump and a conical probe,} {\protect\JournalTitle{Applied Physics Letters}}
	\textbf{110}, 241103 (2017).
	
	\bibitem{lettPRA08}
	C.~F. McCormick, A.~M. Marino, V.~Boyer, and P.~D. Lett, \enquote{Strong
		low-frequency quantum correlations from a four-wave-mixing amplifier,}
	{\protect\JournalTitle{Phys. Rev. A}} \textbf{78}, 043816 (2008).
	
	\bibitem{spiralbandwidthpreprint}
	R.~F. {Offer}, D.~{Stulga}, E.~{Riis}, S.~{Franke-Arnold}, and A.~S. {Arnold},
	\enquote{{Spiral bandwidth of four-wave mixing in Rb vapour},}
	{\protect\JournalTitle{ArXiv e-prints}}  (2018).
	
	\bibitem{Maleev:03}
	I.~D. Maleev and G.~A. Swartzlander, \enquote{Composite optical vortices,}
	{\protect\JournalTitle{J. Opt. Soc. Am. B}} \textbf{20}, 1169--1176 (2003).
	
	\bibitem{PhysRevLett.88.013601}
	G.~Molina-Terriza, J.~P. Torres, and L.~Torner, \enquote{Management of the
		angular momentum of light: Preparation of photons in multidimensional vector
		states of angular momentum,} {\protect\JournalTitle{Phys. Rev. Lett.}}
	\textbf{88}, 013601 (2001).
	
	\bibitem{PhysRevA.67.052313}
	J.~P. Torres, Y.~Deyanova, L.~Torner, and G.~Molina-Terriza,
	\enquote{Preparation of engineered two-photon entangled states for
		multidimensional quantum information,} {\protect\JournalTitle{Phys. Rev. A}}
	\textbf{67}, 052313 (2003).
	
\end{thebibliography}
\end{document}